\documentclass[twoside]{article}
\usepackage{qic}

\usepackage{amsmath}
\usepackage{amssymb}
\usepackage{braket}

\newcommand{\ourtitle}{Limit theorems for the interference terms of discrete-time quantum walks on the line}

\begin{document}
\setlength{\textheight}{8.0truein}    

\runninghead{\ourtitle}
            {T. Machida}

\normalsize\textlineskip
\thispagestyle{empty}
\setcounter{page}{1}



\alphfootnote

\fpage{1}

\centerline{\bf
\ourtitle}
\vspace*{0.37truein}
\centerline{\footnotesize
Takuya Machida}
\vspace*{0.015truein}
\centerline{\footnotesize\it Meiji Institute for Advanced Study of Mathematical Sciences,}
\baselineskip=10pt
\centerline{\footnotesize\it Meiji University, 1-1-1 Higashimita, Tamaku, Kawasaki 214-8571, Japan}
\vspace*{0.225truein}

\vspace*{0.21truein}

\abstracts{
The probability distributions of discrete-time quantum walks have been often investigated, and many interesting properties of them have been discovered.
The probability that the walker can be find at a position is defined by diagonal elements of the density matrix.
On the other hand, although off-diagonal parts of the density matrices have an important role to quantify quantumness, they have not received attention in quantum walks.
We focus on the off-diagonal parts of the density matrices for discrete-time quantum walks on the line and derive limit theorems for them. 
}{}{}

\vspace*{10pt}

\keywords{2-state quantum walk, density matrix, interference term}

\vspace*{1pt}\textlineskip    

\bibliographystyle{qic}

\section{Introduction}
The discrete-time quantum walk (QW) is considered as a quantum analogous of the random walk which expresses the random motion of particles, and many interesting properties of the QWs have been discovered~\cite{AmbainisBachNayakVishwanathWatrous2001,Kempe2003,Kendon2007,Konno2008}.
Although QWs have simple dynamics in analogy with random walks, the behavior of quantum walkers is different from that of random walkers.
We can expect quantum walks to find many applications in quantum computers.
Indeed, compared with classical search algorithms, the corresponding quantum algorithms based on QWs produce exponential speed-up.
The difference between QWs and random walks is seen in probability distributions.
The probability that the quantum walker on the line is found at position $x\in\mathbb{Z}=\left\{0,\pm 1,\pm 2,\ldots\right\}$ is defined by sum of diagonal elements in the density matrix at position $x$.
The probability distributions of the QWs have been analyzed mainly and many limit theorems for them have been proved.
Meanwhile, off-diagonal parts of the density matrices have got little attention in quantum walks and the behavior of them have not clarified.
So, the aim of our paper is to get the characteristics of the off-diagonal parts in QWs. 
Off-diagonal parts of the density matrix, which are often called interference terms, show the distinction between quantum and classical states.
If the particle is in a mixed state, the off-diagonal parts of the density matrix are necessarily null.
That is, the interference terms of the density matrix for classical state are vanished.
When the state of particles is not classical, they can become non-zero values.
The off-diagonal parts of the density matrix have an important role when we characterize quantumness of particles.
In quantum mechanics they are the essential values to express decoherence, which has been also investigated in QWs, both numerically and analytically~\cite{Kendon2007,AbalDonangeloSeveroSiri2008,AnnabestaniAkhtarshenasAbolhassani2010,SrikanthBanerjeeChandrashekar2010,Zhang2008}.

Diagonal parts of the density matrix of the QW after long-time were obtained in a limit theorem of the probability distribution by Konno\,\cite{Konno2002,Konno2005}, while the direct study for the off-diagonal parts of the density matrices in QWs has not been done.
So, the author was motivated to clarify the long-time asymptotic behavior of the off-diagonal parts. 
Instead, we have analyzed the von Neumann entropy to quantify the entanglement, which is defined by density matrices~\cite{AbalSiriRomanelliDonangelo2006,AnnabestaniAbolhasaniAbal2010,CarneiroLooXuGirerdKendonKnight2005,IdeKonnoMachida2011,LiuPetulante2010}.
Carneiro et al.~\cite{CarneiroLooXuGirerdKendonKnight2005} numerically computed the von Neumann entropy of QWs.
In Abal et al.~\cite{AbalSiriRomanelliDonangelo2006} and Annabestani et al.~\cite{AnnabestaniAbolhasaniAbal2010}, analytical results were obtained.
Liu and Petulante~\cite{LiuPetulante2010} and Ide et al.~\cite{IdeKonnoMachida2011} derived limit theorems for the von Neumann entropy.

The present paper is organized as follows.
In Sect.~2, we introduce the notations of 2-state QWs on the line and give the time evolution rule and probability distribution of the QWs.
In Sect.~3, we present two limit theorems for interference terms of density matrices as our main result.
Section 4 is devoted to proofs of the limit theorems.
Summary is mentioned in the last section.
In appendix A, we show limit theorems for 3-state Grover walks on the line.


\section{Definition of the 2-state QW on the line}
\label{sec:definition}
In this section we define the 2-state QW on the line.
Let $\ket{x}$ ($x\in\mathbb{Z}=\left\{0,\pm 1,\pm2,\ldots\right\}$) be infinite components vectors which denote the position of the walker.
Here,  $x$-th component of $\ket{x}$ is 1 and the other is 0.
Let $\ket{\psi_{t}(x)} \in \mathbb{C}^2$ be the amplitude of the walker at position $x$ at time $t \in\left\{0,1,2,\ldots\right\}$, where $\mathbb{C}$ is the set of complex numbers.
The walk at time $t$ is expressed by
\begin{equation}
 \ket{\Psi_t}=\sum_{x\in\mathbb{Z}}\ket{x}\otimes\ket{\psi_{t}(x)}.
\end{equation}
The time evolution of the walk is described by the following unitary matrix:
\begin{equation}
  U=\left[\begin{array}{cc}
    u_{00} & u_{01} \\ u_{10} & u_{11}
	 \end{array}\right],
\end{equation}
where $u_{j_1j_2}\,\in\mathbb{C}\, (j_1,j_2\in\left\{0,1\right\})$.
In the present paper, we take the components of $U$ as $u_{00}=-u_{11}=\cos\theta, u_{01}=u_{10}=\sin\theta$ or $u_{00}=u_{11}=\cos\theta, -u_{01}=u_{10}=\sin\theta$ with $\theta\in (0,2\pi)\, (\theta\neq \pi/2, \pi, 3\pi/2)$.
Then the evolution is determined by
\begin{equation}
 \ket{\psi_{t+1}(x)}=P\ket{\psi_t(x+1)}+Q\ket{\psi_t(x-1)},\label{eq:te}
\end{equation}
where $P=\ket{0}\bra{0}U, Q=\ket{1}\bra{1}U$ and
\begin{equation}
 \ket{0}=\left[\begin{array}{c}
  1\\0
       \end{array}\right],\,
 \ket{1}=\left[\begin{array}{c}
  0\\1
       \end{array}\right].
\end{equation}
\noindent The probability that the quantum walker $X_t$ is at position $x$ at time $t$, $\mathbb{P}(X_t=x)$, is defined by
\begin{equation}
 \mathbb{P}(X_t=x)=\sum_{j=0}^{1}\bra{j}\rho_t(x)\ket{j},\label{eq:prob}
\end{equation}
where $\rho_t(x)=\ket{\psi_t(x)}\bra{\psi_t(x)}$ are the density matrices at position $x$ at time $t$.

\clearpage

\section{Main results}
\label{sec:result}

We will show two limit theorems for the off-diagonal parts of the density matrices $\rho_t(x)$.
A limit theorem for the diagonal elements $\bra{j}\rho_t(x)\ket{j}\,(j\in\left\{0,1\right\})$, that is probability distribution, has been given already by Konno~\cite{Konno2002,Konno2005} and Grimmett et al.~\cite{GrimmettJansonScudo2004}. 
Konno~\cite{Konno2002,Konno2005} obtained the limit theorem for the probability distribution of $X_t/t$ as $t\to\infty$ by using path counting method.
Grimmett et al.~\cite{GrimmettJansonScudo2004} also computed it by the Fourier analysis.
From their results, for the 2-state QWs starting from the origin with $\ket{\psi_0(0)}=\alpha\ket{0}+\beta\ket{1}$ and $|\alpha|^2+|\beta|^2=1$, we see the following limit theorem.

\begin{figure}[h]
 \begin{center}
  \includegraphics[scale=0.3]{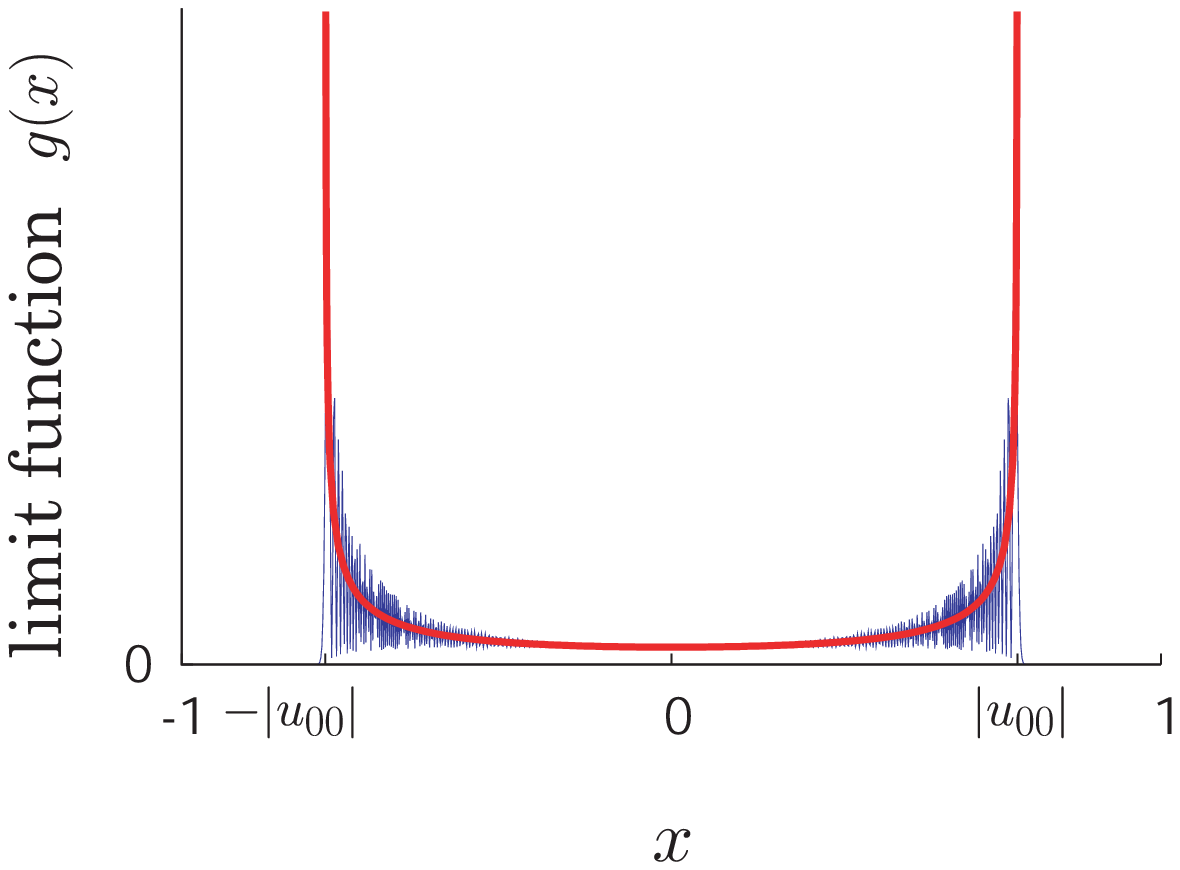}
 \end{center}
 \fcaption{The limit function $g(x)$ (thick line) and probability distribution of $X_t/t$ at time $t=1000$ (thin line) with $u_{00}=u_{01}=u_{10}=-u_{11}=1/\sqrt{2},\alpha=1/\sqrt{2},\beta=i/\sqrt{2}$}
 \label{fig:2-state_limprob}
\end{figure}

\noindent For $u_{00}u_{01}u_{10}u_{11}\neq 0$ and $r=0,1,2,\ldots$, we have
\begin{equation}
 \lim_{t\to\infty}\sum_{x\in\mathbb{Z}}\left(\frac{x}{t}\right)^r\sum_{j=0}^{1}\bra{j}\rho_t(x)\ket{j}
 =\lim_{t\to\infty}\mathbb{E}\left[\left(\frac{X_t}{t}\right)^{r}\right]=\int_{-\infty}^{\infty}y^r g(y)\,dy,
\end{equation}
where
\begin{align}
 g(x)=&\frac{\sqrt{1-|u_{00}|^2}}{\pi(1-x^2)\sqrt{|u_{00}|^2-x^2}}I_{(-|u_{00}|,|u_{00}|)}(x)\nonumber\\
 &\times\left\{1-\left(|\alpha|^2-|\beta|^2+\frac{\alpha u_{00}\overline{\beta u_{01}}+\overline{\alpha u_{00}}\beta u_{01}}{|u_{00}|^2}\right)x\right\},
\end{align}
$I_A(x)=1$ if $x\in A$, $I_A(x)=0$ if $x\notin A$ and $\mathbb{E}(Y)$ means the expected value of $Y$.
We compare the limit function $g(x)$ with probability distribution of the 2-state QW at time $t=1000$ in Fig.~\ref{fig:2-state_limprob}.

In the present paper, we focus on the off-diagonal parts $\bra{0}\rho_t(x)\ket{1}$ in the density matrices $\rho_t(x)$, which are often called the interference terms in quantum mechanics, and derive limit theorems for them.
At first, when we consider the 2-state quantum walks starting with any initial state, we can obtain the following relation between the interference terms and the moments of $X_t/t$.

\begin{lemma}
For $r=0,1,2,\ldots$, we have
\begin{equation}
   \lim_{t\to\infty}\sum_{x\in\mathbb{Z}}\left(\frac{x}{t}\right)^r \Re(\bra{0}\rho_t(x)\ket{1})=(\det U)\frac{s}{2c}\,\lim_{t\to\infty}\mathbb{E}\left[\left(\frac{X_t}{t}\right)^{r+1}\right],\label{eq:2-state_th1}
\end{equation}
where $c=\cos\theta, s=\sin\theta$ and $\Re(z)$ denotes the real part of $z\in\mathbb{C}$.
\label{th:2-state_1}
\end{lemma}
\clearpage

\noindent Moreover, if we assume that the 2-state quantum walks starting from the origin with $\ket{\psi_0(0)}=\alpha\ket{0}+\beta\ket{1}$ and $|\alpha|^2+|\beta|^2=1$, the limit theorem is given as follows.

\begin{theorem}
For $r=0,1,2,\ldots$, we see
\begin{equation}
  \lim_{t\to\infty}\sum_{x\in\mathbb{Z}}\left(\frac{x}{t}\right)^r \bra{0}\rho_t(x)\ket{1}=\int_{-\infty}^{\infty}\,y^r\left\{f^{(R)}(y)+if^{(I)}(y)\right\}\,dy,
\end{equation}
where
\begin{align}
 f^{(R)}(x)=&(\det U)\,\frac{s}{2c}x\cdot\frac{|s|}{\pi(1-x^2)\sqrt{c^2-x^2}}\nonumber\\
 &\times\left[1-\left\{|\alpha|^2-|\beta|^2-(\det U)\frac{s}{c}(\alpha\overline{\beta}+\overline{\alpha}\beta)\right\}x\right]I_{(-|c|,|c|)}(x),\\[5mm]
 f^{(I)}(x)=&\frac{|s|\Im(\alpha\overline{\beta})}{\pi c^2}\cdot\frac{\sqrt{c^2-x^2}}{1-x^2}I_{(-|c|,|c|)}(x),
\end{align}
and $\Im(z)$ is the imaginary part of $z\in\mathbb{C}$.
\label{th:2-state_2}
\end{theorem}
\vspace{1mm}

\begin{figure}[h]
\begin{center}
\begin{minipage}{50mm}
 \begin{center}
  \includegraphics[scale=0.3]{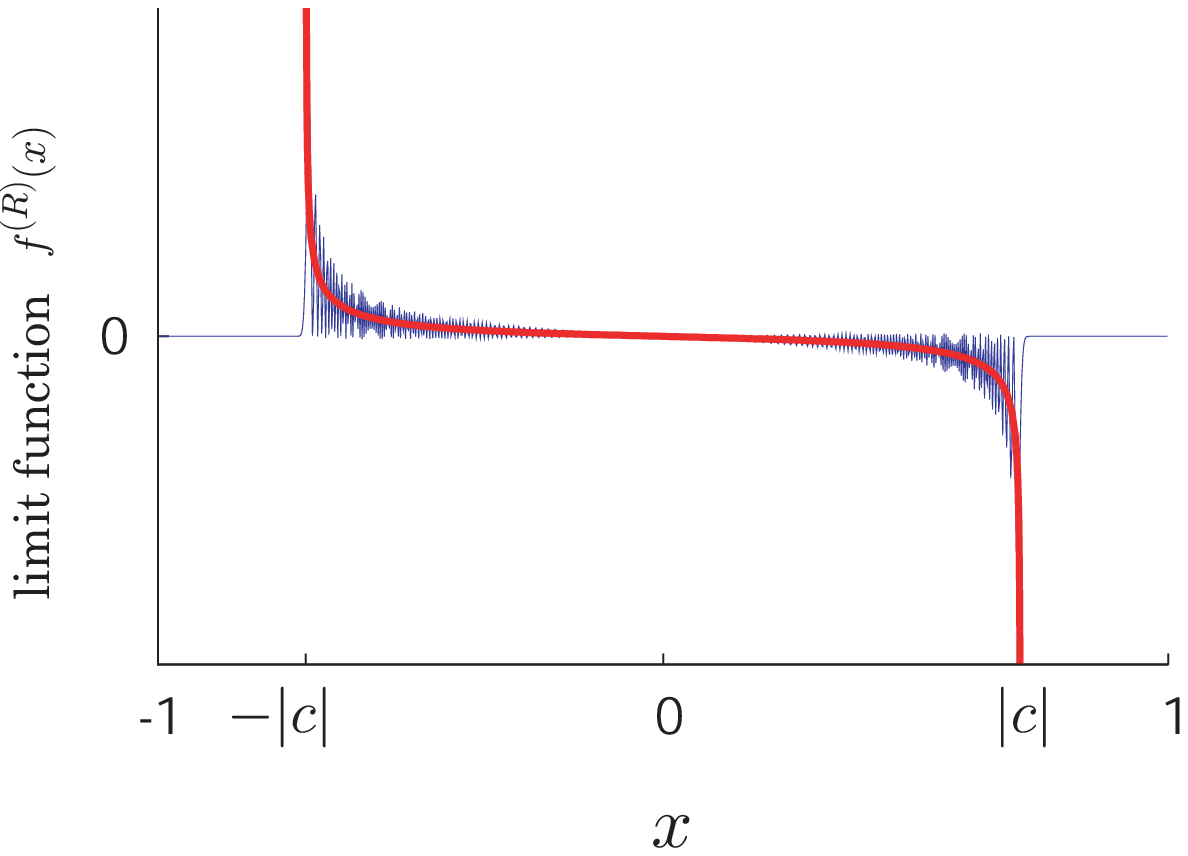}\\
  {\small (a) $f^{(R)}(x), \Re(\bra{0}\rho_t(x)\ket{1})$}
 \end{center}
\end{minipage}
\begin{minipage}{50mm}
 \begin{center}
  \includegraphics[scale=0.3]{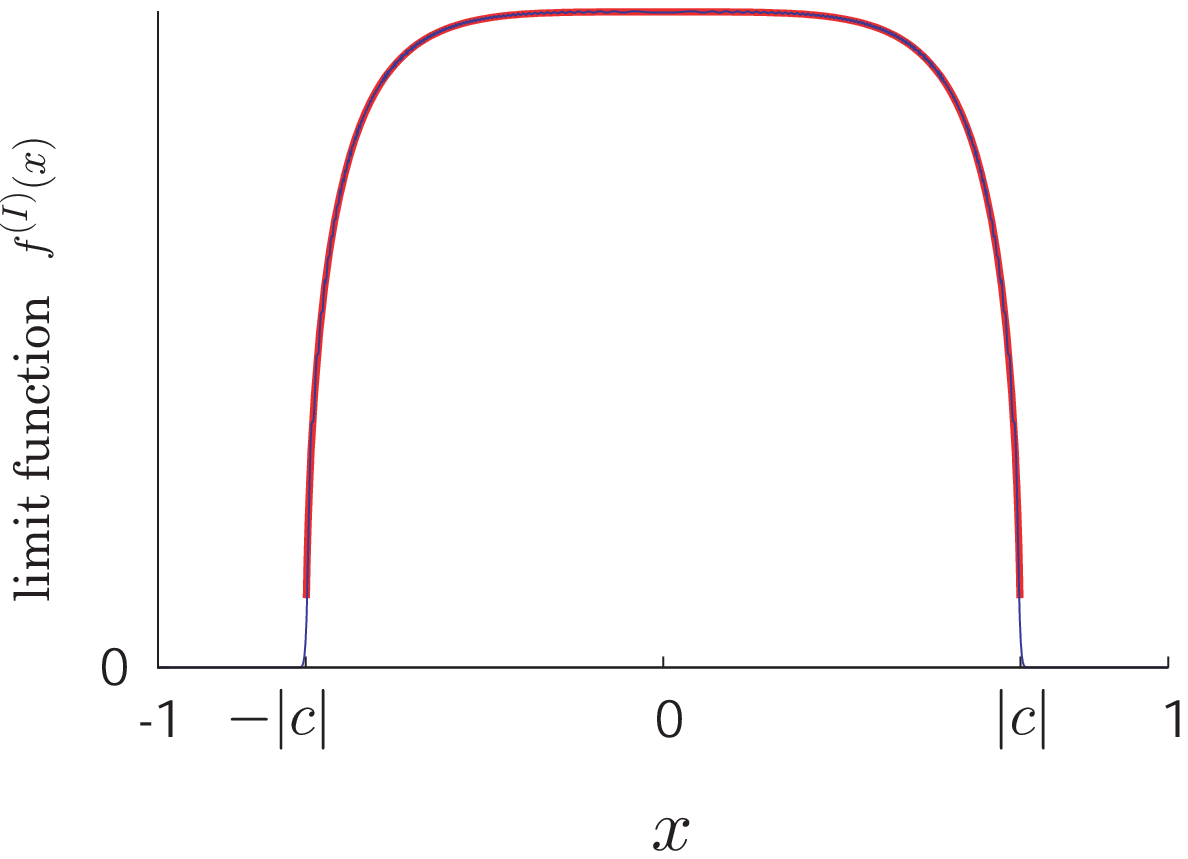}\\
  {\small (b) $f^{(I)}(x), \Im(\bra{0}\rho_t(x)\ket{1})$}
 \end{center}
\end{minipage}
\vspace{5mm}
\fcaption{The limit function $f^{(R)}(x), f^{(I)}(x)$ (thick line) and $\Re(\bra{0}\rho_t(x)\ket{1}), \Im(\bra{0}\rho_t(x)\ket{1})$ at time $t=1000$ (thin line) with $u_{00}=-u_{11}=c=1/\sqrt{2}, u_{01}=u_{10}=s=1/\sqrt{2},\alpha=1/\sqrt{2},\beta=i/\sqrt{2}$}
\label{fig:2-state_interface_term}
\end{center}
\end{figure}

\noindent Figure~\ref{fig:2-state_interface_term} shows a comparison of the limit functions $f^{(R)}(x), f^{(I)}(x)$ and the off-diagonal parts of the density matrices $\rho_t(x)$.

\section{Proof of our results}

\subsection{Proof of Lemma~\ref{th:2-state_1}}

Our proof is based on the Fourier analysis introduced by Grimmett et al.~\cite{GrimmettJansonScudo2004}.
The Fourier transform $\ket{\hat{\Psi}_{t}(k)}\,(k\in\left[-\pi,\pi\right))$ of $\ket{\psi_t(x)}$ is given by
\begin{equation}
 \ket{\hat{\Psi}_{t}(k)}=\sum_{x\in\mathbb{Z}} e^{-ikx}\ket{\psi_t(x)}.\label{eq:ft}
\end{equation}
By the inverse Fourier transform, we have
\begin{equation}
 \ket{\psi_t(x)}=\int_{-\pi}^{\pi}e^{ikx}\ket{\hat\Psi_{t}(k)}\,\frac{dk}{2\pi}.
\end{equation}
From Eqs. (\ref{eq:te}) and (\ref{eq:ft}), the time evolution of $\ket{\hat{\Psi}_{t}(k)}$ becomes
\begin{equation}
 \ket{\hat{\Psi}_{t+1}(k)}= \hat U(k)\ket{\hat{\Psi}_{t}(k)},\label{eq:te_1step}
\end{equation}
where $\hat U(k)=R(k)U$ and $R(k)=e^{ik}\ket{0}\bra{0}+e^{-ik}\ket{1}\bra{1}$.
Equation~(\ref{eq:te_1step}) implies
\begin{equation}
 \ket{\hat{\Psi}_{t}(k)}=\hat U(k)^t \ket{\hat\Psi_{0}(k)}.\label{eq:ft_timet}
\end{equation}
The eigenvalues $\lambda_j(k)\,(j\in\left\{1,2\right\})$ of the unitary matrix $\hat U(k)$ are
\begin{equation}
 \lambda_j(k)=\left\{\begin{array}{ll}
		ic\sin k-(-1)^j\sqrt{1-c^2\sin^2 k}& (u_{00}=-u_{11}=c, u_{01}=u_{10}=s),\\[2mm]
		c\cos k-i(-1)^j \sqrt{1-c^2\cos^2 k}& (u_{00}=u_{11}=c, -u_{01}=u_{10}=s),
		      \end{array}\right.
\end{equation}
and the eigenvectors corresponding to $\lambda_j(k)$ take the following form
\begin{equation}
 \left\{\begin{array}{ll}
		\left[\begin{array}{c}
		 se^{ik}\\ -c\cos k-(-1)^j\sqrt{1-c^2\sin^2 k}
		      \end{array}\right]& (u_{00}=-u_{11}=c, u_{01}=u_{10}=s),\\[4mm]
		\left[\begin{array}{c}
		 se^{ik}\\ i\left\{c\sin k+(-1)^j\sqrt{1-c^2\cos^2 k}\right\}
		      \end{array}\right]& (u_{00}=u_{11}=c, -u_{01}=u_{10}=s).
		      \end{array}\right.
\end{equation}
By the eigenvalues $\lambda_j(k)$ and the normalized eigenvectors $\ket{v_j(k)}$, Eq.~(\ref{eq:ft_timet}) can be rewritten as
\begin{align}
 \ket{\hat{\Psi}_{t}(k)}=\hat U(k)^t \sum_{j=1}^{2}\braket{v_j(k)|\hat\Psi_0(k)}\ket{v_j(k)}=\sum_{j=1}^2 \lambda_j(k)^t \braket{v_j(k)|\hat\Psi_0(k)}\ket{v_j(k)}.
\end{align}

\noindent From now, we will prove Lemma~\ref{th:2-state_1}.
At first, in order to express the left-side hand of Eq.~(\ref{eq:2-state_th1}) by the Fourier transform, we compute as follows:
\begin{align}
 &\sum_{x\in\mathbb{Z}}x^r\Re(\bra{0}\rho_t(x)\ket{1})=\frac{1}{2}\sum_{x\in\mathbb{Z}}x^r\left\{\bra{0}\rho_t(x)\ket{1}+\bra{1}\rho_t(x)\ket{0}\right\}\nonumber\\
 =&\frac{1}{2}\int_{-\pi}^{\pi}\bra{\hat\Psi_{t}(k)}J^{+}\left(D^r\ket{\hat\Psi_{t}(k)}\right)\,\frac{dk}{2\pi}\nonumber\\
 =&\frac{(t)_r}{2}\int_{-\pi}^{\pi}\Biggl\{\,\sum_{j=1}^{2}h_j(k)^r|\braket{v_j(k)|\hat\Psi_0(k)}|^2\braket{v_j(k)|J^{+}|v_j(k)}\nonumber\\
 &+(\overline{\lambda_1(k)}\lambda_2(k))^t h_2(k)^r\braket{v_2(k)|\hat\Psi_0(k)}\braket{v_1(k)|J^{+}|v_2(k)}\overline{\braket{v_1(k)|\hat\Psi_0(k)}}\nonumber\\
 &+(\lambda_1(k)\overline{\lambda_2(k)})^t h_1(k)^r\braket{v_1(k)|\hat\Psi_0(k)}\braket{v_2(k)|J^{+}|v_1(k)}\overline{\braket{v_2(k)|\hat\Psi_0(k)}}\Biggr\}\,\frac{dk}{2\pi}\nonumber\\
 &+O(t^{r-1}),
\end{align}
where $J^{+}=\ket{0}\bra{1}+\ket{1}\bra{0}$, $h_j(k)=D\lambda_j(k)/\lambda_j(k)\,(j\in\left\{1,2\right\})$, $D=i(d/dk)$ and $(t)_r=t(t-1)\times\cdots\times(t-r+1)$.
By using the Riemann-Lebesgue lemma, we have
\begin{align}
 \lim_{t\to\infty}\sum_{x\in\mathbb{Z}}\left(\frac{x}{t}\right)^r\Re(\bra{0}\rho_t(x)\ket{1})
  =&\frac{1}{2}\int_{-\pi}^{\pi}\left\{\,\sum_{j=1}^{2}h_j(k)^r|\braket{v_j(k)|\hat\Psi_0(k)}|^2\braket{v_j(k)|J^{+}|v_j(k)}\right\}\,\frac{dk}{2\pi}.
\end{align}
Since we easily find $\braket{v_j(k)|J^{+}|v_j(k)}=(\det\,U)\frac{s}{c}h_j(k)$,
\begin{align}
 \lim_{t\to\infty}\sum_{x\in\mathbb{Z}}\left(\frac{x}{t}\right)^r\Re(\bra{0}\rho_t(x)\ket{1})
  =&(\det\,U)\frac{s}{2c}\int_{-\pi}^{\pi}\left\{\,\sum_{j=1}^{2}h_j(k)^{r+1}|\braket{v_j(k)|\hat\Psi_0(k)}|^2\right\}\,\frac{dk}{2\pi}\nonumber\\
  =&(\det\,U)\frac{s}{2c}\,\lim_{t\to\infty}\mathbb{E}\left[\left(\frac{X_t}{t}\right)^{r+1}\right].
\end{align}
\begin{flushright}
 $\Box$
\end{flushright}

\subsection{Proof of Theorem~\ref{th:2-state_2}}

\noindent The real part $f^{(R)}(x)$ can be obtained from Lemma~\ref{th:2-state_1} and the result in Konno~\cite{Konno2002,Konno2005} as follows:
\begin{align}
 \lim_{t\to\infty}\sum_{x\in\mathbb{Z}}\left(\frac{x}{t}\right)^r \Re(\bra{0}\rho_t(x)\ket{1})
  =&(\det U)\frac{s}{2c}\,\int_{-\infty}^{\infty} y^{r+1}\,\frac{|s|}{\pi(1-y^2)\sqrt{c^2-y^2}}\nonumber\\
 &\times\left[1-\left\{|\alpha|^2-|\beta|^2-(\det U)\frac{s}{c}(\alpha\overline{\beta}+\overline{\alpha}\beta)\right\}y\right]I_{(-|c|,|c|)}(y)\,dy\nonumber\\
 =&\int_{-\infty}^{\infty} y^{r}\,(\det U)\,\frac{s}{2c}y\cdot\frac{|s|}{\pi(1-y^2)\sqrt{c^2-y^2}}\nonumber\\
 &\times\left[1-\left\{|\alpha|^2-|\beta|^2-(\det U)\frac{s}{c}(\alpha\overline{\beta}+\overline{\alpha}\beta)\right\}y\right]I_{(-|c|,|c|)}(y)\,dy.\label{eq:re_limit_2state}
\end{align}
Next, we calculate the imaginary part $f^{(I)}(x)$.
In a similar fashion as the proof of Lemma~\ref{th:2-state_1},
\begin{align}
 \lim_{t\to\infty}\sum_{x\in\mathbb{Z}}\left(\frac{x}{t}\right)^r\Im(\bra{0}\rho_t(x)\ket{1})
  =&\frac{1}{2i}\int_{-\pi}^{\pi}\left\{\,\sum_{j=1}^{2}h_j(k)^r|\braket{v_j(k)|\psi_0(0)}|^2\braket{v_j(k)|J^{-}|v_j(k)}\right\}\,\frac{dk}{2\pi},
\end{align}
where $\Im(z)$ denotes the imaginary part of $z\in\mathbb{C}$ and $J^{-}=\ket{1}\bra{0}-\ket{0}\bra{1}$.
Note that the initial state becomes
$\ket{\hat\Psi_0(k)}=\ket{\psi_0(0)}=\alpha\ket{0}+\beta\ket{1}$.
After putting $h_j(k)=y\,(j\in\left\{1,2\right\})$, we get
\begin{equation}
 \lim_{t\to\infty}\sum_{x\in\mathbb{Z}}\left(\frac{x}{t}\right)^r\Im(\bra{0}\rho_t(x)\ket{1})
  =\int_{-\infty}^{\infty} y^r\,\frac{|s|\Im(\alpha\overline{\beta})}{\pi c^2}\cdot\frac{\sqrt{c^2-y^2}}{1-y^2}I_{(-|c|,|c|)}(y)\,dy.\label{eq:im_limit_2state}
\end{equation}
Combining Eqs. (\ref{eq:re_limit_2state}) and (\ref{eq:im_limit_2state}), we complete the proof of Theorem~\ref{th:2-state_2}.
\begin{flushright}
 $\Box$
\end{flushright}

\section{Summary}
In this section, we mention discussion and conclusion for our results.
In the present paper, we investigated off-diagonal parts in the density matrices of discrete-time 2-state QWs on the line.
In Ide et al.\,\cite{IdeKonnoMachida2011}, an entropy of entanglement $-\text{Tr}\left\{\rho_t^c\log_2(\rho_t^c)\right\}$ for the QW was analyzed in the limit $t\to\infty$, where $\rho_t^c=\sum_{x\in\mathbb{Z}}\rho_t(x)$ is the density matrix over the whole system at time $t$ (see p. 857 in their paper).
That is, the values $\rho_t(x)$ quantify the entanglement of the QW.
The diagonal parts of the density matrix $\rho_t(x)$ after long-time were directly computed as a limit theorem of the probability distribution in Konno\,\cite{Konno2002,Konno2005}.  
In this paper, we gave a limit theorem to the off-diagonal parts. 
Off-diagonal parts of the density matrix are called interference terms and they are important values to quantify decoherence of quantum particles.
In Sect.~\ref{sec:result} we gave two limit theorems for the interference terms of density matrices at position $x\in\mathbb{Z}$ in 2-state QWs.
One is the limit theorem for the QW starting with any initial state, the other is the limit theorem when the walker starts from the origin. 
Lemma~\ref{th:2-state_1} showed the relation between the interference terms of the QW starting with any initial state and the $r$-th moments ($r=0,1,2,\ldots$) of $X_t/t$.
From Theorem~\ref{th:2-state_2}, if the walker starts from the origin, the limit function which denotes the imaginary part of the interference terms always becomes an even function regardless of the initial state at the origin.
One of the interesting problems is to get the relations between decoherent process and QWs, that is, expression of transition from quantum state to classical state by QWs.
Chisaki et al.~\cite{ChisakiKonnoSegawaShikano2011} studied crossover from QW to classical random walk.
So, it would be interesting to discuss the relation between our results and their results.
If we can construct the decoherent process by QWs, innovative investigations may start in quantum mechanics.

\nonumsection{Acknowledgements}
\noindent The author is grateful to Norio Konno and Etsuo Segawa for useful comment and also to Joe Yuichiro Wakano and the Meiji University Global COE Program ``Formation and Development of Mathematical Sciences Based on Modeling and Analysis'' for the support.

\nonumsection{References}

\begin{thebibliography}{10}

\bibitem{AmbainisBachNayakVishwanathWatrous2001}
A. Ambainis, E. Bach, A. Nayak, A. Vishwanath and J. Watrous
\newblock (2001), \textit{One-dimensional quantum walks}, Proceedings of the 33rd annual ACM symposium on Theory of
  computing, ACM, pp.  37--49.

\bibitem{Kempe2003}
J. Kempe
\newblock (2003), \textit{Quantum random walks: an introductory overview},
  Contemporary Physics, 44, pp.  307--327.

\bibitem{Kendon2007}
V. Kendon
\newblock (2007), \textit{Decoherence in quantum walks -- a review}, Mathematical
  Structures in Computer Science, 17, pp.  1169--1220.

\bibitem{Konno2008}
N. Konno
\newblock (2008),
\newblock \textit{Quantum Walks}, Lecture Notes in Mathematics,
  Springer-Verlag
\newblock Heidelberg, 1954, pp.  309--452.

\bibitem{AbalDonangeloSeveroSiri2008}
G. Abal, R. Donangelo, F. Severo and R. Siri
\newblock (2008), \textit{Decoherent quantum walks driven by a generic coin
  operation}, Physica A: Statistical Mechanics and its Applications, 387,
  pp.  335--345.

\bibitem{AnnabestaniAkhtarshenasAbolhassani2010}
M. Annabestani, S.J. Akhtarshenas and M.R. Abolhassani
\newblock (2010), \textit{Decoherence in a one-dimensional quantum walk},
  Phys. Rev. A, 81,  032321.

\bibitem{SrikanthBanerjeeChandrashekar2010}
R. Srikanth, S. Banerjee and C.M. Chandrashekar
\newblock (2010), \textit{Quantumness in a decoherent quantum walk using
  measurement-induced disturbance}, Phys. Rev. A, 81,  062123.

\bibitem{Zhang2008}
K. Zhang
\newblock (2008), \textit{Limiting distribution of decoherent quantum random
  walks}, Phys. Rev. A, 77,  062302.

\bibitem{AbalSiriRomanelliDonangelo2006}
G. Abal, R. Siri, A. Romanelli and R. Donangelo
\newblock (2006), \textit{Quantum walk on the line: Entanglement and nonlocal
  initial conditions}, Phys. Rev. A, 73,  042302.

\bibitem{AnnabestaniAbolhasaniAbal2010}
M. Annabestani, M.R. Abolhasani and G. Abal
\newblock (2010), \textit{Asymptotic entanglement in 2D quantum walks}, Journal
  of Physics A: Mathematical and Theoretical, 43,  075301.

\bibitem{CarneiroLooXuGirerdKendonKnight2005}
I. Carneiro, M. Loo, X. Xu, M. Girerd, V. Kendon and P.L. Knight
\newblock (2005), \textit{Entanglement in coined quantum walks on regular
  graphs}, New Journal of Physics, 7,  156.

\bibitem{IdeKonnoMachida2011}
Y. Ide, N. Konno and T. Machida
\newblock (2011), \textit{Entanglement for discrete-time quantum walks on the
  line}, Quantum Information and Computation, 11, pp.  855--866.

\bibitem{LiuPetulante2010}
C. Liu and N. Petulante
\newblock (2010), \textit{On the von Neumann entropy of certain quantum walks
  subject to decoherence}, Mathematical Structures in Computer Science, 20,
  pp.  1099--1115.

\bibitem{Konno2002}
N. Konno
\newblock (2002), \textit{Quantum random walks in one dimension},
 Quantum Information Processing,
  1, pp.  345--354.

\bibitem{Konno2005}
N. Konno
\newblock (2005), \textit{A new type of limit theorems for the one-dimensional
  quantum random walk}, Journal of the Mathematical Society of Japan, 57, pp.
  1179--1195.

\bibitem{GrimmettJansonScudo2004}
G. Grimmett, S. Janson and P.F. Scudo
\newblock (2004), \textit{Weak limits for quantum random walks}, Phys.
  Rev. E, 69,  026119.

\bibitem{InuiKonno2005}
N. Inui and N. Konno
\newblock (2005), \textit{Localization of multi-state quantum walk in one
  dimension}, Physica A: Statistical Mechanics and its Applications, 353, pp.
  133--144.

\bibitem{ChisakiKonnoSegawaShikano2011}
K. Chisaki, N. Konno, E. Segawa and Y. Shikano
\newblock (2011), \textit{Crossovers induced by discrete-time quantum walks}, Quantum Information and Computation, 11, pp.
  741--760.

\end{thebibliography}

\appendix

\begin{figure}[h]
 \begin{center}
  \includegraphics[scale=0.29]{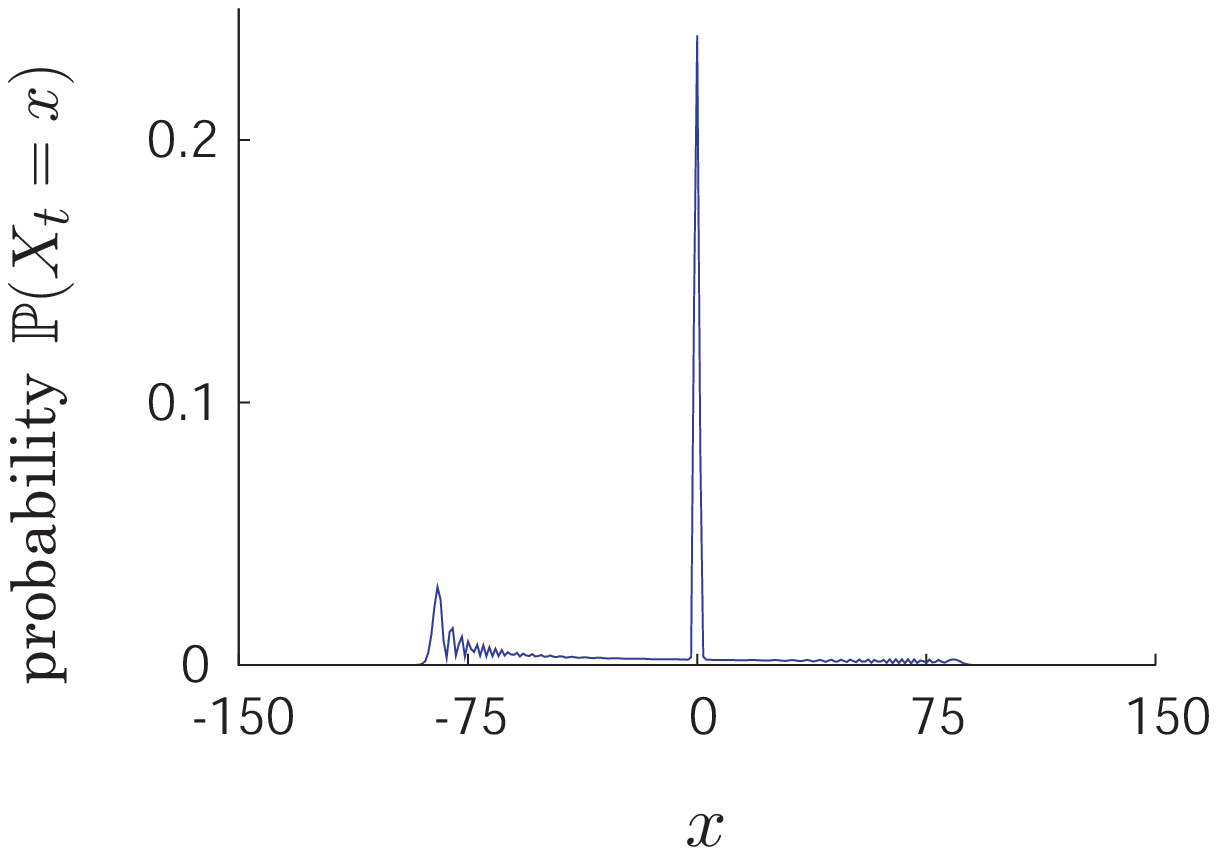}
 \end{center}
 \fcaption{Probability distribution of the 3-state Grover walk at time $t=150$ starting from the origin with the initial state $\ket{\psi_0(0)}=1/\sqrt{3}\,\ket{0}+i/\sqrt{3}\,\ket{1}+i/\sqrt{3}\,\ket{2}$}
 \label{fig:3-state}
\end{figure}

\noindent In this appendix, we consider the 3-state Grover walk on the line and will show two limit theorems for the off-diagonal parts of the density matrices.
The difference between the 2-state QWs defined in Sect.~\ref{sec:definition} and the 3-state Grover walk is whether localization occurs or not.
The probability distribution of the 2-state QWs doesn't localized, while the 3-state Grover walk is known as one of the models in which localization can occur.
The 3-state Grover walk on the line is defined as follows.
The amplitudes of the walker at position $x\in\mathbb{Z}$ at time $t \in\left\{0,1,2,\ldots\right\}$ are given by $\ket{\psi_{t}(x)} \in \mathbb{C}^3$.
The time evolution of the walk is determined by the Grover matrix $G=\frac{2}{3}\sum_{j_1=0}^2\sum_{j_2=0}^2 \ket{j_1}\bra{j_2}-\sum_{j=0}^2\ket{j}\bra{j}$ with
\begin{equation}
 \ket{0}=\left[\begin{array}{c}
	  1\\0\\0
	       \end{array}\right],\,
 \ket{1}=\left[\begin{array}{c}
	  0\\1\\0
	       \end{array}\right],\,
 \ket{2}=\left[\begin{array}{c}
	  0\\0\\1
	       \end{array}\right].
\end{equation}
The amplitudes $\ket{\psi_t(x)}$ evolve by the following rule:
\begin{equation}
 \ket{\psi_{t+1}(x)}=P_0\ket{\psi_t(x+1)}+P_1\ket{\psi_t(x)}+P_2\ket{\psi_t(x-1)},\label{eq:teGW}
\end{equation}
where $P_0=\ket{0}\bra{0}G, P_1=\ket{1}\bra{1}G, P_2=\ket{2}\bra{2}G$.
Inui et al.~\cite{InuiKonno2005} focused on the probability distribution $\mathbb{P}(X_t=x)=\sum_{j=0}^{2}\bra{j}\rho_t(x)\ket{j}$ and got a long-time limit theorem for $\sum_{x\in\mathbb{Z}}\left(\frac{x}{t}\right)^r\sum_{j=0}^{2}\bra{j}\rho_t(x)\ket{j}=\mathbb{E}\left[(X_t/t)^r\right]$ (see Eq.~(16) in~\cite{InuiKonno2005}).
Figure~\ref{fig:3-state} shows probability distribution of the 3-state Grover walk at time $t=150$ with $\ket{\psi_0(0)}=1/\sqrt{3}\,\ket{0}+i/\sqrt{3}\,\ket{1}+i/\sqrt{3}\,\ket{2}$.

\noindent On the other hand, we derive limit theorems for the interference terms $\bra{j_1}\rho_t(x)\ket{j_2} ((j_1,j_2)\in\left\{(0,1),(0,2),(1,2)\right\})$.
The proofs of theorems are given as well as the ones of both Lemma~\ref{th:2-state_1} and Theorem~\ref{th:2-state_2}.
So, we will omit them.\\

For the 3-state Grover walk starting with any initial state, we can obtain the relations between the interference terms and the moments of $X_t/t$.

\begin{lemma}
As $t\to\infty$, the summations of $(x/t)^r \Re(\bra{j_1}\rho_t(x)\ket{j_2}) \,(r=0,1,2,\ldots)$ all over $x\in\mathbb{Z}$ are calculated as follows:
\begin{enumerate}
 \item For $r=0$, we have
       \begin{align}
	&\lim_{t\to\infty}\sum_{x\in\mathbb{Z}}\Re(\bra{j_1}\rho_t(x)\ket{j_2})\nonumber\\
	&=\left\{\begin{array}{ll}
	   \Delta_{01}^{(R)}+\frac{1}{2}\lim_{t\to\infty}\mathbb{E}\left[\frac{X_t}{t}\right]-\frac{1}{2}\lim_{t\to\infty}\mathbb{E}\left[\left(\frac{X_t}{t}\right)^2\right]& (j_1=0,j_2=1),\\
		  \Delta_{02}^{(R)}+\frac{1}{4}-\frac{5}{4}\lim_{t\to\infty}\mathbb{E}\left[\left(\frac{X_t}{t}\right)^2\right]& (j_1=0,j_2=2),\\
		  \Delta_{12}^{(R)}-\frac{1}{2}\lim_{t\to\infty}\mathbb{E}\left[\frac{X_t}{t}\right]-\frac{1}{2}\lim_{t\to\infty}\mathbb{E}\left[\left(\frac{X_t}{t}\right)^2\right]& (j_1=1,j_2=2),		  
		 \end{array}\right.\label{eq:th3r0}
       \end{align}
       where
       \begin{equation}
	\Delta_{j_1j_2}^{(R)}=\left\{\begin{array}{ll}
			     \int_{-\pi}^\pi\,\frac{dk}{2\pi}\,\frac{1}{2}|\braket{v(k)|\hat\Psi_0(k)}|^2\braket{v(k)|J_{j_1j_2}^{+}|v(k)}& ((j_1,j_2)=(0,1),(1,2)),\\[2mm]
				    \int_{-\pi}^\pi\,\frac{dk}{2\pi}\,\left\{\frac{1}{2}|\braket{v(k)|\hat\Psi_0(k)}|^2\braket{v(k)|J_{j_1j_2}^{+}|v(k)}\right.& \\
				      \qquad\qquad\left.-\frac{1}{4}|\braket{v(k)|\hat\Psi_0(k)}|^2\right\}& ((j_1,j_2)=(0,2)),\\
				    
				   \end{array}\right.
       \end{equation}
       and $\ket{\hat\Psi_0(k)}=\sum_{x\in\mathbb{Z}}e^{-ikx}\ket{\psi_0(x)}$ and $J_{j_1j_2}^{+}=\ket{j_1}\bra{j_2}+\ket{j_2}\bra{j_1}$.
       The vector $\ket{v(k)}$ is the normalized eigenvector corresponding to the eigenvalue $\lambda=1$ of the unitary matrix $R(k)G$ with
       $R(k)=e^{ik}\ket{0}\bra{0}+\ket{1}\bra{1}+e^{-ik}\ket{2}\bra{2}$:
       \begin{align}
	\ket{v(k)}=&\sqrt{\frac{2}{5+\cos k}}\left[\begin{array}{c}
					      1\\ (1+e^{-ik})/2\\ e^{-ik}
						   \end{array}\right].
       \end{align}

 \item For $r=1,2,\ldots$, we have
       \begin{align}
	&\lim_{t\to\infty}\sum_{x\in\mathbb{Z}}\left(\frac{x}{t}\right)^r \Re(\bra{j_1}\rho_t(x)\ket{j_2})\\
	&=\left\{\begin{array}{ll}
	   \frac{1}{2}\lim_{t\to\infty}\mathbb{E}\left[\left(\frac{X_t}{t}\right)^{r+1}\right]-\frac{1}{2}\lim_{t\to\infty}\mathbb{E}\left[\left(\frac{X_t}{t}\right)^{r+2}\right]& (j_1=0,j_2=1),\\[2mm]
		  \frac{1}{4}\lim_{t\to\infty}\mathbb{E}\left[\left(\frac{X_t}{t}\right)^{r}\right]-\frac{5}{4}\lim_{t\to\infty}\mathbb{E}\left[\left(\frac{X_t}{t}\right)^{r+2}\right]& (j_1=0,j_2=2),\\[2mm]
		  -\frac{1}{2}\lim_{t\to\infty}\mathbb{E}\left[\left(\frac{X_t}{t}\right)^{r+1}\right]-\frac{1}{2}\lim_{t\to\infty}\mathbb{E}\left[\left(\frac{X_t}{t}\right)^{r+2}\right]& (j_1=1,j_2=2).
		 \end{array}\right.\label{eq:th3r}
       \end{align}
\end{enumerate}
\label{th:3-state_1}
\end{lemma}
\vspace{5mm}

\noindent Particularly, when the 3-state Grover walk starts from the origin with $\ket{\psi_0(0)}=\alpha\ket{0}+\beta\ket{1}+\gamma\ket{2}$ and $|\alpha|^2+|\beta|^2+|\gamma|^2=1$, we can get the specific limit theorem for the interference terms.

\begin{theorem}
For $r=0,1,2,\ldots$, we see
\begin{equation}
 \lim_{t\to\infty}\sum_{x\in\mathbb{Z}}\left(\frac{x}{t}\right)^r \bra{j_1}\rho_t(x)\ket{j_2}=\int_{-\infty}^{\infty}\,y^r\left\{f_{j_1j_2}^{(R)}(y)+if_{j_1j_2}^{(I)}(y)\right\}\,dy,
\end{equation}
where
\begin{align}
 f_{j_1j_2}^{(R)}(x)=&\left\{\begin{array}{ll}
		       \Delta_{01}^{(R)}\delta_0(x)+\frac{\sqrt{2}(c_0+c_1x+c_2x^2)}{4\pi\sqrt{1-3x^2}}\frac{x}{1+x}I_{(-\frac{1}{\sqrt{3}},\frac{1}{\sqrt{3}})}(x)&(j_1=0, j_2=1),\\[3mm]
			      \Delta_{02}^{(R)}\delta_0(x)+\frac{\sqrt{2}(c_0+c_1x+c_2x^2)}{4\pi\sqrt{1-3x^2}}\frac{1-5x^2}{2(1-x^2)}I_{(-\frac{1}{\sqrt{3}},\frac{1}{\sqrt{3}})}(x)&(j_1=0, j_2=2),\\[3mm]
			       \Delta_{12}^{(R)}\delta_0(x)-\frac{\sqrt{2}(c_0+c_1x+c_2x^2)}{4\pi\sqrt{1-3x^2}}\frac{x}{1-x}I_{(-\frac{1}{\sqrt{3}},\frac{1}{\sqrt{3}})}(x)&(j_1=1, j_2=2),
			     \end{array}\right.\\[2mm]
 f_{j_1j_2}^{(I)}(x)=&\left\{\begin{array}{ll}
		       \Delta_{01}^{(I)}\delta_0(x)+\frac{(d_0+d_1 x)\sqrt{1-3x^2}}{\pi}\frac{1}{\sqrt{2}(1+x)}I_{(-\frac{1}{\sqrt{3}},\frac{1}{\sqrt{3}})}(x)&(j_1=0, j_2=1),\\[3mm]
			\Delta_{02}^{(I)}\delta_0(x)-\frac{(d_0+d_1 x)\sqrt{1-3x^2}}{\pi}\frac{\sqrt{2}x}{1-x^2}I_{(-\frac{1}{\sqrt{3}},\frac{1}{\sqrt{3}})}(x)&(j_1=0, j_2=2),\\[3mm]
			\Delta_{12}^{(I)}\delta_0(x)-\frac{(d_0+d_1 x)\sqrt{1-3x^2}}{\pi}\frac{1}{\sqrt{2}(1-x)}I_{(-\frac{1}{\sqrt{3}},\frac{1}{\sqrt{3}})}(x)&(j_1=1, j_2=2),
			     \end{array}\right.
\end{align}
and $\delta_0(x)$ denotes Dirac's $\delta$-function at the origin.
The coefficients are computed as follows:
\begin{align}
 \Delta_{j_1j_2}^{(R)}=&\left\{\begin{array}{ll}
		  \frac{\sqrt{6}}{36}\left(\Bigl|\alpha+\frac{\beta}{2}\Bigr|^2+\Bigl|\gamma+\frac{\beta}{2}\Bigr|^2\right)&\\
 +\left(1-\frac{29\sqrt{6}}{72}\right)\Re\left\{(2\alpha+\beta)(2\overline{\gamma}+\overline{\beta})\right\}&(j_1=0,j_2=1),\\[3mm]
			  -\frac{\sqrt{6}}{72}\left(\Bigl|\alpha+\frac{\beta}{2}\Bigr|^2+\Bigl|\gamma+\frac{\beta}{2}\Bigr|^2\right)&\\
 +\left(2-\frac{115\sqrt{6}}{144}\right)\Re\left\{(2\alpha+\beta)(2\overline{\gamma}+\overline{\beta})\right\}&(j_1=0,j_2=2),\\[3mm]
			  \frac{\sqrt{6}}{36}\left(\Bigl|\alpha+\frac{\beta}{2}\Bigr|^2+\Bigl|\gamma+\frac{\beta}{2}\Bigr|^2\right)&\\
 +\left(1-\frac{29\sqrt{6}}{72}\right)\Re\left\{(2\alpha+\beta)(2\overline{\gamma}+\overline{\beta})\right\}&(j_1=1,j_2=2),
			\end{array}\right.\\[3mm]
 \Delta_{j_1j_2}^{(I)}=&\left\{\begin{array}{ll}
		  \left(2-\frac{5\sqrt{6}}{6}\right)\Im(\overline{\alpha}\beta+\overline{\beta}\gamma+2\gamma\overline{\alpha})&(j_1=0,j_2=1),\\[3mm]
			  \left(4-\frac{5\sqrt{6}}{3}\right)\Im(\overline{\alpha}\beta+\overline{\beta}\gamma+2\gamma\overline{\alpha})&(j_1=0,j_2=2),\\[3mm]
			  \left(2-\frac{5\sqrt{6}}{6}\right)\Im(\overline{\alpha}\beta+\overline{\beta}\gamma+2\gamma\overline{\alpha})&(j_1=1,j_2=2),
			  \end{array}\right.
\end{align}
and
\begin{align}
 c_0&=|\alpha+\gamma|^2+2|\beta|^2,\\
 c_1&=2\left(-|\alpha-\beta|^2+|\gamma-\beta|^2\right),\\
 c_2&=|\alpha-\gamma|^2-2\Re\left\{(2\alpha+\beta)(2\overline{\gamma}+\overline{\beta}))\right\},\\
 d_0&=\Im\left\{(\alpha+\gamma)\overline{\beta}\right\},\\
 d_1&=\Im\left\{\overline{\alpha}(\beta+\gamma)+(\overline{\alpha}+\overline{\beta})\gamma\right\}.
\end{align}
\label{th:3-state_2}
\end{theorem}

\begin{figure}[h]
 \begin{center}
  \begin{minipage}{45mm}
   \begin{center}
    \includegraphics[scale=0.3]{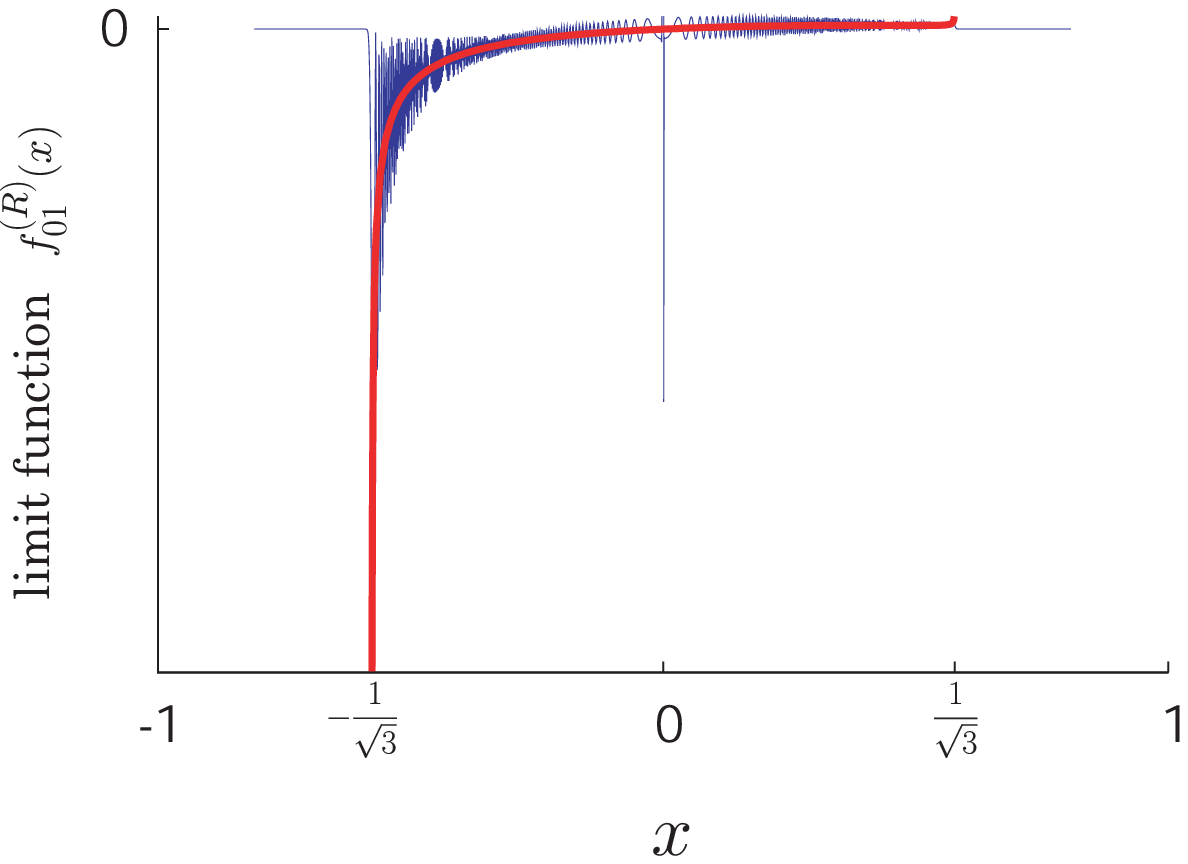}\\
    {\small (a) $(j_1,j_2)=(0,1)$}
   \end{center}
  \end{minipage}
  \begin{minipage}{45mm}
   \begin{center}
    \includegraphics[scale=0.3]{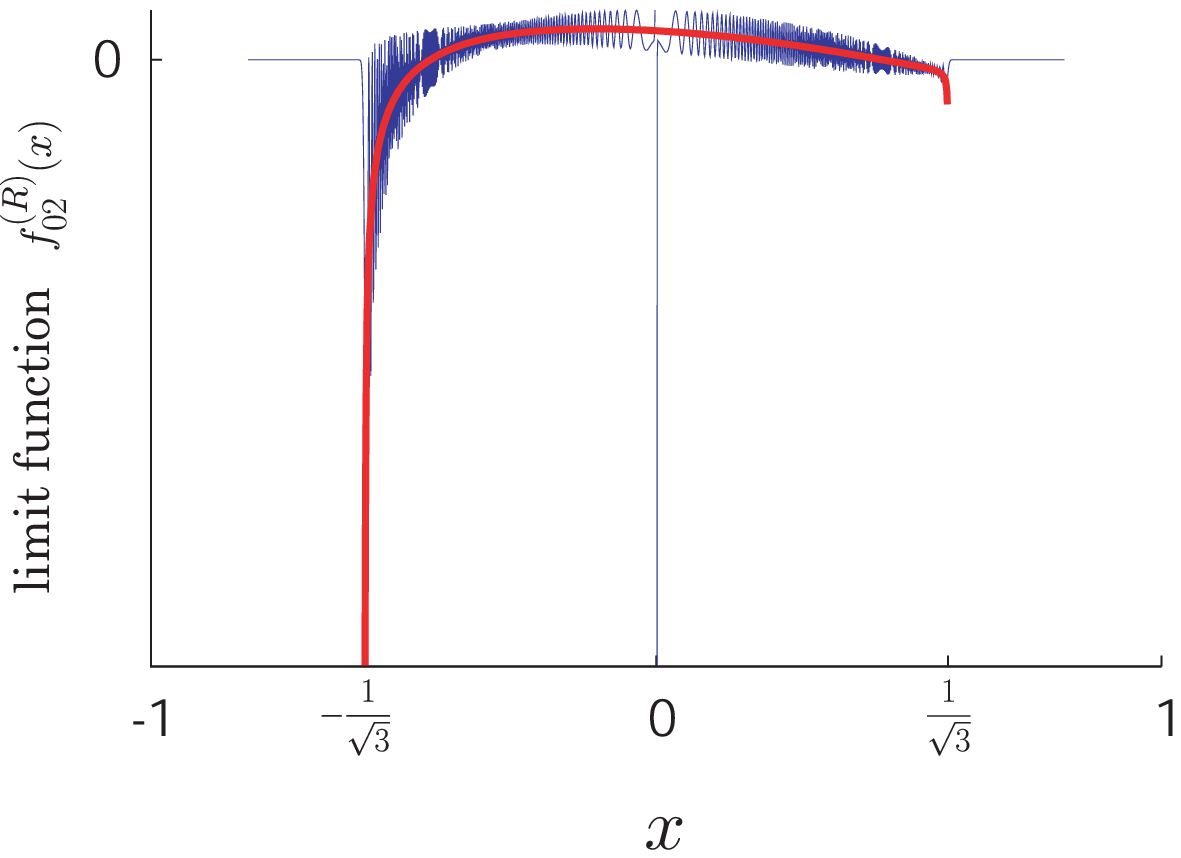}\\
    {\small (b) $(j_1,j_2)=(0,2)$}
   \end{center}
  \end{minipage}
  \begin{minipage}{45mm}
   \begin{center}
    \includegraphics[scale=0.3]{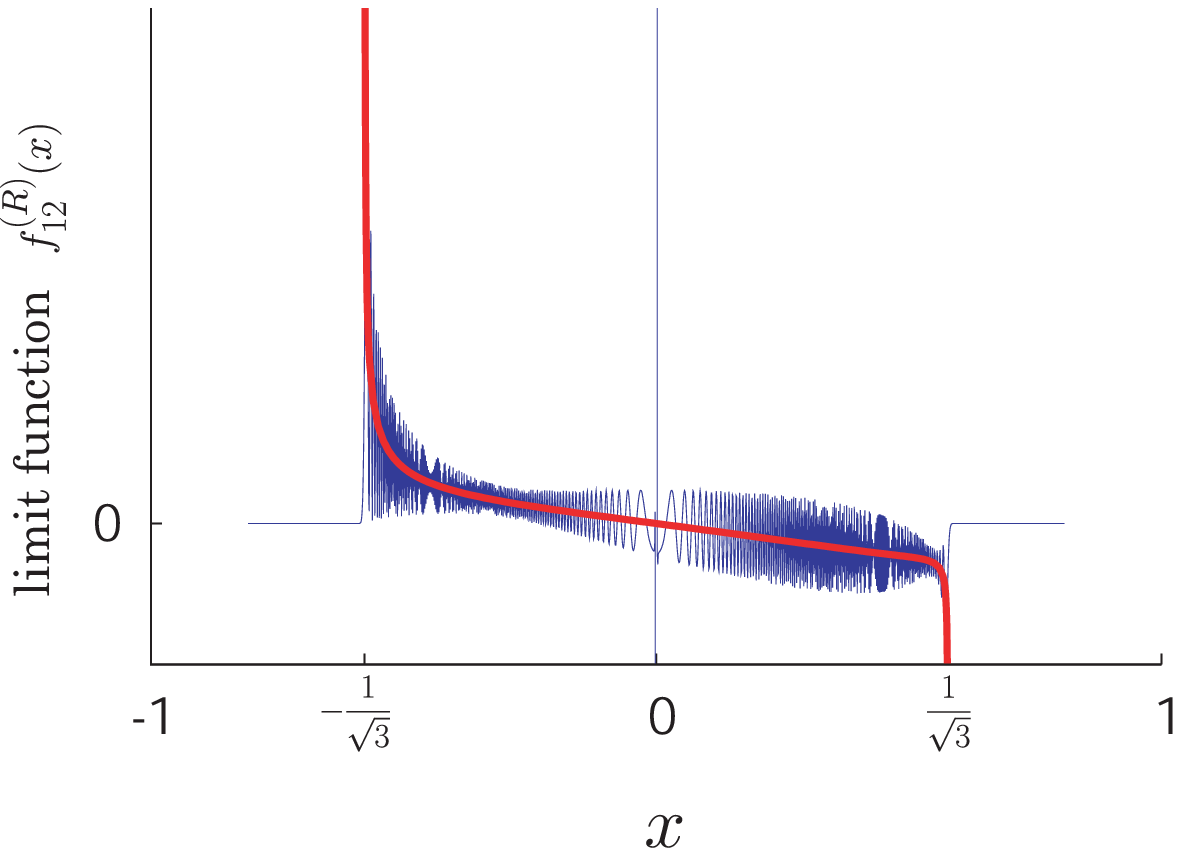}\\
    {\small (c) $(j_1,j_2)=(1,2)$}
   \end{center}
  \end{minipage}
  \vspace{5mm}
  \fcaption{The limit function $f_{j_1j_2}^{(R)}(x)$ (thick line) and $\Re(\bra{j_1}\rho_t(x)\ket{j_2})$ at time $t=1000$ (thin line) with $\alpha=1/\sqrt{3},\,\beta=\gamma=i/\sqrt{3}$}
  \label{fig:3-state_interface_term_re}
  \end{center}
\end{figure}

\begin{figure}[h]
 \begin{center}
    \begin{minipage}{45mm}
  \begin{center}
   \includegraphics[scale=0.3]{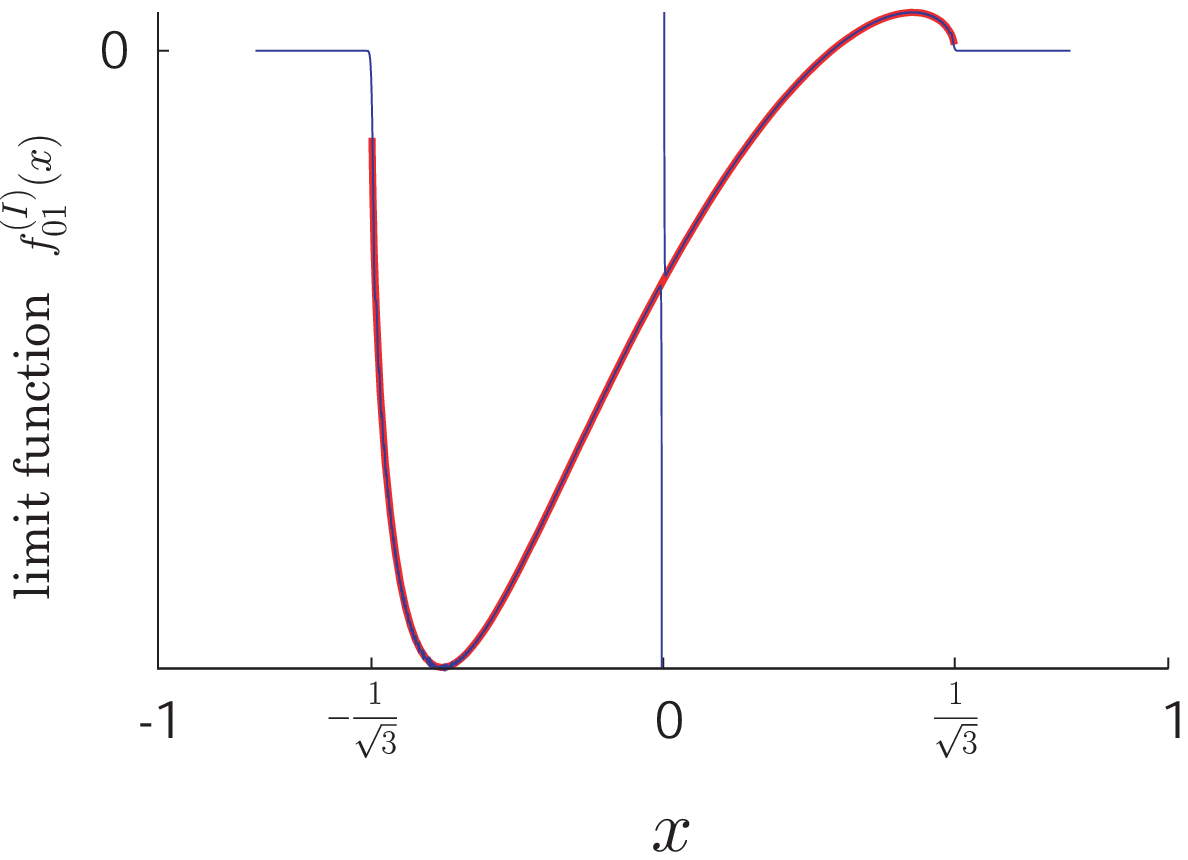}\\
   {\small (a) $(j_1,j_2)=(0,1)$}
  \end{center}
  \end{minipage}
  \begin{minipage}{45mm}
  \begin{center}
   \includegraphics[scale=0.3]{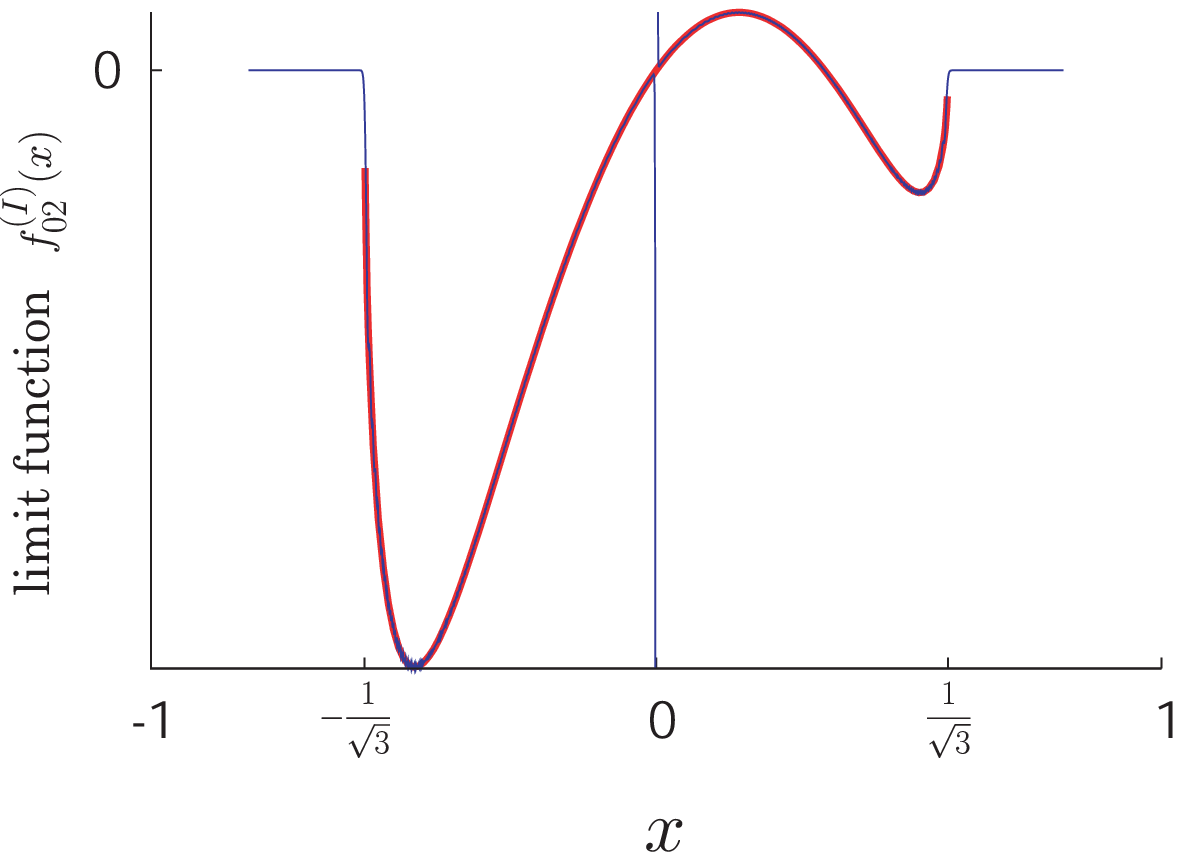}\\
   {\small (b) $(j_1,j_2)=(0,2)$}
  \end{center}
  \end{minipage}
    \begin{minipage}{45mm}
  \begin{center}
   \includegraphics[scale=0.3]{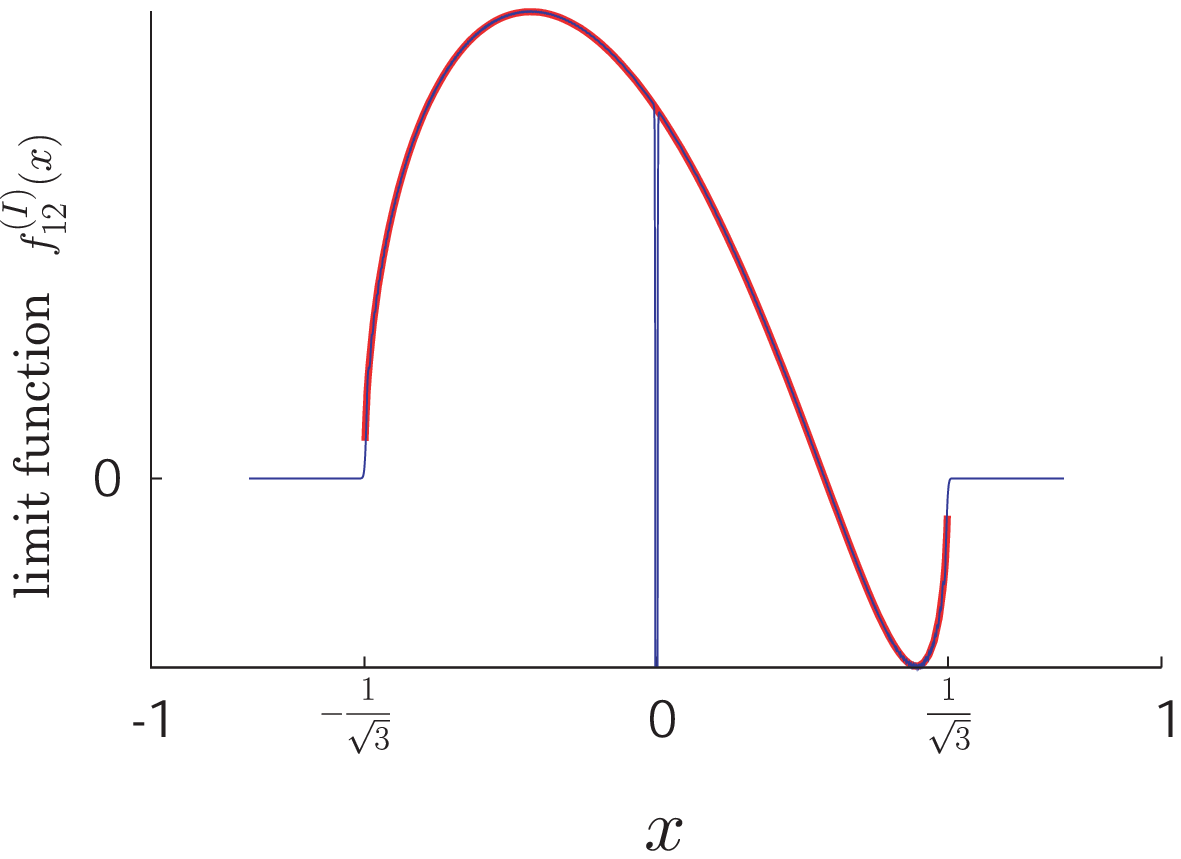}\\
   {\small (c) $(j_1,j_2)=(1,2)$}
  \end{center}
  \end{minipage}
  \vspace{5mm}
  \fcaption{The limit function $f_{j_1j_2}^{(I)}(x)$ (thick line) and $\Im(\bra{j_1}\rho_t(x)\ket{j_2})$ at time $t=1000$ (thin line) with $\alpha=1/\sqrt{3},\,\beta=\gamma=i/\sqrt{3}$}
  \label{fig:3-state_interface_term_im}
 \end{center}
\end{figure}

\noindent We should note $\Delta_{01}^{(R)}=\Delta_{12}^{(R)}, \Delta_{01}^{(I)}=\Delta_{12}^{(I)}$.
When Lemma~\ref{th:3-state_1} and Theorem~\ref{th:3-state_2} are proved, we consider $J_{j_1j_2}^{+}=\ket{j_1}\bra{j_2}+\ket{j_2}\bra{j_1}$ (resp. $J_{j_1j_2}^{-}=\ket{j_2}\bra{j_1}-\ket{j_1}\bra{j_2}$) instead of $J^{+}=\ket{0}\bra{1}+\ket{1}\bra{0}$ (resp. $J^{-}=\ket{1}\bra{0}-\ket{0}\bra{1}$) in the proofs of both Lemma~\ref{th:2-state_1} and Theorem~\ref{th:2-state_2}.
Examples of Theorem~\ref{th:3-state_2} are presented in Figs.~\ref{fig:3-state_interface_term_re} and \ref{fig:3-state_interface_term_im}. 
\clearpage

\end{document}